\documentclass[11pt,a4paper]{article}
\usepackage[utf8]{inputenc}
\usepackage{amsmath,amssymb}
\usepackage{graphicx}
\usepackage{booktabs}
\usepackage[margin=2.5cm]{geometry}
\usepackage[font=small,labelfont=bf]{caption}
\usepackage[expansion=false]{microtype}
\usepackage[numbers,sort&compress]{natbib}
\usepackage[colorlinks=true,linkcolor=blue,citecolor=blue,urlcolor=blue]{hyperref}

\newcommand{\ad}{a^{\dagger}}
\newcommand{\ket}[1]{|#1\rangle}
\newcommand{\avg}[1]{\langle #1\rangle}

\title{Fragmented and symmetry-restored condensates in a three-site Bose ring:\\ exchange versus umklapp}

\author{Afif Siddiki\\[4pt]
\small Atlas University Vocational School, \.{I}stanbul, Turkey\\
\small \href{mailto:afif.siddiki@atlas.edu.tr}{afif.siddiki@atlas.edu.tr}}

\date{}

\begin{document}
\maketitle

\begin{abstract}
We study $N$ bosons on a three-site ring with nearest-neighbour hopping, density-assisted hopping of the opposite sign, and an on-site interaction, written equivalently as three momentum modes $q=0,\pm1$ with a channel-dependent quartic vertex.
As the density-assisted term grows it inverts the band, and exact diagonalisation up to $N=400$ shows a first-order ground-state transition from the $q=0$ condensate to a state with $\avg{n_{+1}}=\avg{n_{-1}}\simeq N/2$; the avoided-crossing gap closes faster than exponentially with $N$.
In its symmetric eigenstates this state is fragmented in the Penrose-Onsager sense, and at leading order in $N$ it has the energy of a Gross-Pitaevskii (GP) standing wave, as expected for a few-mode mean-field system.
We distinguish two cases by switching off umklapp scattering, which is available because $2\cdot(+1)\equiv-1$ on three sites, and by varying the on-site interaction.
Without umklapp, $n_{+1}-n_{-1}$ is conserved, the eigenstates form a rotor tower whose lowest member is a fragmented Fock state, and the transition leads to fragmentation exactly when the $\pm1$ interaction is attractive, as predicted by Nozi\`eres' exchange argument with the sign of the interaction reversed; otherwise the system forms a single current-carrying condensate.
With umklapp, the standing wave is pinned to three positions, the three momentum sectors become degenerate exponentially in $N$, and coherent admixture of the $q=0$ mode stabilises the standing wave over a wider range; the exact ground state is then a number-squeezed, momentum-projected standing-wave condensate rather than a robustly fragmented state.
\end{abstract}

\section{Introduction}

Whether interacting bosons condense into a single one-particle state or fragment into several macroscopically occupied states is decided, according to Nozi\`eres~\cite{Nozieres1982,Nozieres1995}, by exchange energy: for repulsive interactions, distributing particles over two degenerate modes costs exchange energy, so the condensate stays in one mode.
Fragmentation is defined through the one-body density matrix~\cite{PenroseOnsager1956,Leggett2001}, a state being fragmented when more than one eigenvalue is of order $N$.
Fragmented ground states nevertheless occur, for example in spin-1 gases~\cite{LawPuBigelow1998,HoYip2000}, in double wells~\cite{SpekkensSipe1999} and for attractive bosons on a ring~\cite{UedaLeggett1999,Kanamoto2003}; Mueller \textit{et al.}~\cite{Mueller2006} reviewed the conditions under which they arise and stressed that they are fragile against perturbations that break the protecting symmetry.

A useful distinction in this context is between a fragmented state proper, in which the occupations of the macroscopically occupied modes are sharp, and a symmetry-restored single condensate, i.e.\ a mean-field condensate in a symmetry-breaking orbital projected onto a symmetry sector.
Both give the same one-body density matrix in a symmetric eigenstate.
They differ in their number fluctuations and in whether a mean-field (GP) state captures the energy.
For attractive bosons on a ring, for instance, the uniform condensate gives way to a bright soliton; the many-body ground state is the translationally symmetric superposition of solitons, and the mean-field transition at $\gamma=-1/2$ becomes a smooth crossover with large number fluctuations~\cite{Kanamoto2003}.

Here we analyse a minimal model in which both kinds of state occur, and in which the mechanism that produces them can be switched.
It is a three-site Bose-Hubbard ring with density-assisted (correlated) hopping~\cite{Dutta2015,Meinert2016} whose sign is opposite to that of the single-particle hopping.
At low density the bosons condense at $k=0$; when the density-assisted term dominates, the effective hopping changes sign, and on the frustrated triangle the band minimum moves to the degenerate pair $k=\pm2\pi/3$.
The question is then how the bosons occupy this degenerate pair: in one of them, forming a current-carrying condensate, or in both, forming a standing wave or a fragmented state.
The model descends from a finite, discrete anharmonic-phonon model studied by the author in 1999~\cite{thesis,preprint2006}.
In that work the question was posed whether exchange correlations, following Nozi\`eres, decide condensation in a canonical ensemble~\cite{thesis}, but the approximate (Gaussian and grand-canonical) methods used there could not answer it.
In particular, the exact canonical treatment presented here does not support the condensation temperature reported in Refs.~\cite{thesis,preprint2006}; for a few-mode system, the canonical thermodynamic functions are smooth.

Our results are as follows.
(i) At fixed $N$ the $q=0$ condensate undergoes a first-order transition into a state with $\avg{n_{+1}}=\avg{n_{-1}}\simeq N/2$ and zero total momentum (Sec.~\ref{sec:transition}).
(ii) For a fixed number of modes with interaction $\propto1/N$ the leading-order energy is given by the coherent-state (Hartree) functional~\cite{LewinNamRougerie2014,LNSS2015}; we use this to obtain the critical coupling and the $q=0$ admixture, and check that exact diagonalisation up to $N=400$ converges to them as $1/N$ (Sec.~\ref{sec:largeN}).
(iii) In both versions of the model the fragmented state is the symmetric eigenstate of a GP standing wave; what distinguishes them is the symmetry that the standing wave breaks, a continuous one without umklapp and a discrete $\mathbb{Z}_3$ with umklapp, and this shows up directly in the tower of low-lying states (Sec.~\ref{sec:nature}).
(iv) Without umklapp, fragmentation versus a single current-carrying condensate is decided exactly by the sign of the exchange energy; with umklapp, coherent coupling to the $q=0$ mode stabilises the standing wave even for a repulsive $\pm1$ interaction (Sec.~\ref{sec:mech}).

\section{Model and methods}\label{sec:model}

\subsection{Hamiltonian}
We consider $N$ bosons in three modes labelled by the crystal momentum $q\in\{0,+1,-1\}$, defined modulo 3, with
\begin{equation}
H=\sum_q\varepsilon_q n_q+\frac{\lambda}{24N}\sum_{q_0+q_1\equiv q_2+q_3}M(q_0,q_1,q_2,q_3)\,
\ad_{q_0}\ad_{q_1}a_{q_2}a_{q_3},
\label{eq:H}
\end{equation}
where $\varepsilon_q=\hbar\omega|q|$ (linear dispersion, inherited from the phonon origin of the model), $\equiv$ denotes equality modulo 3, and the coupling is scaled by $1/N$ so that the interaction energy per particle stays finite.
Energies are in units of $\hbar\omega$.
The reference vertex is
\begin{equation}
M(q_0,q_1,q_2,q_3)=\sum_{i=0}^{3}\cos\frac{2\pi q_i}{3}+c ,
\label{eq:M}
\end{equation}
where the constant shift $c$ (zero unless stated) is used to test the mechanisms.
The vertex takes only four distinct values, one for each scattering channel:
\begin{equation}
\begin{aligned}
&U_0=4+c\ \ (\text{all four }q=0), && U_1=c-2\ \ (\text{all four }q=\pm1),\\
&U_{01}=1+c\ \ (\text{two }0,\ \text{two }\pm1), && U_{\rm um}=c-\tfrac12\ \ (\text{one }0,\ \text{three }\pm1).
\end{aligned}
\label{eq:channels}
\end{equation}
For $c=0$ the interaction is repulsive in the $q=0$ channel and attractive in the $\pm1$ channel.
The last channel contains only umklapp processes such as $\ad_{+1}\ad_{+1}a_0a_{-1}$, for which momentum is conserved modulo 3 but not as an integer ($+2\equiv-1$).
All other terms conserve the integer momentum $L=n_{+1}-n_{-1}$.
Setting $U_{\rm um}=0$ defines the ``no-umklapp'' model, which we use to isolate the two mechanisms.
A momentum-independent vertex, $M=V_0$, is the situation of Nozi\`eres' argument and serves as a control.

\subsection{Real-space form}
With $a_q=3^{-1/2}\sum_x e^{-2\pi iqx/3}b_x$ for sites $x=0,1,2$, Eq.~(\ref{eq:H}) is identical (we checked the operator identity numerically to $10^{-14}$) to
\begin{equation}
H=\frac{2N}{3}-\frac13\sum_x\left(b^\dagger_xb_{x+1}+{\rm h.c.}\right)
+\frac{\lambda}{8N}\sum_x\Big[c\,n_x(n_x-1)+\sum_{\delta=\pm1}\left(b^\dagger_x n_x b_{x+\delta}+{\rm h.c.}\right)\Big].
\label{eq:Hreal}
\end{equation}
The linear dispersion $\varepsilon_q=|q|$ on three sites is thus ordinary nearest-neighbour hopping, $\varepsilon_q=\tfrac23[1-\cos(2\pi q/3)]$; the constant $c$ is an on-site Hubbard interaction; and the cosine part of the vertex is density-assisted hopping with amplitude $\lambda/(8N)>0$, opposite in sign to the single-particle hopping $-1/3$.
Density-assisted hopping arises in non-standard Hubbard models and has been engineered with ultracold atoms by Floquet modulation~\cite{Dutta2015,Meinert2016}.
A momentum-independent vertex, $M=V_0$, corresponds to a pure on-site interaction.
The ``no-umklapp'' model is a theoretical device: removing the umklapp terms makes the interaction non-local in real space, but it isolates the exchange mechanism.

\subsection{Symmetries and methods}
$H$ conserves $N$, the crystal momentum $P=L \bmod 3$, and the reflection $n_{+1}\leftrightarrow n_{-1}$.
Without umklapp, $L$ itself is conserved.
Because $P$ is conserved, the one-body density matrix $\avg{\ad_qa_{q'}}$ of any eigenstate is diagonal in $q$, and its eigenvalues are the occupations $\avg{n_q}$.
We quantify fragmentation by $F=2\min(\avg{n_{+1}},\avg{n_{-1}})/N$.

We diagonalise $H$ in each $P$ block of the fixed-$N$ Fock space, whose dimension is about $\binom{N+2}{2}/3$ (26\,867 for $N=400$), using sparse Lanczos diagonalisation.
Basis states are stored as integers whose digits are the occupations (Appendix~\ref{app:encoding}).
The critical coupling $\lambda_c$ is defined by $\avg{n_0}/N=1/2$ in the ground state and located by bisection.

For large $N$ we use the coherent-state (GP) energy per particle,
\begin{equation}
e[\psi]=\sum_q\varepsilon_q|\psi_q|^2+\frac{\lambda}{24}\sum_{q_0+q_1\equiv q_2+q_3}M(q_0,q_1,q_2,q_3)\,\psi^*_{q_0}\psi^*_{q_1}\psi_{q_2}\psi_{q_3},
\qquad\sum_q|\psi_q|^2=1,
\label{eq:GP}
\end{equation}
minimised over the three complex amplitudes.
For a fixed number of modes and interaction $\propto1/N$, the minimum of Eq.~(\ref{eq:GP}) gives the ground-state energy per particle to leading order~\cite{LewinNamRougerie2014}, with corrections of order $1/N$ described by Bogoliubov theory~\cite{LNSS2015}.
A state with $|\psi_{+1}|=|\psi_{-1}|\neq0$ is a standing wave that breaks the translational (momentum) symmetry.

\section{First-order transition}\label{sec:transition}

\begin{figure}[t]
\centering
\includegraphics[width=\textwidth]{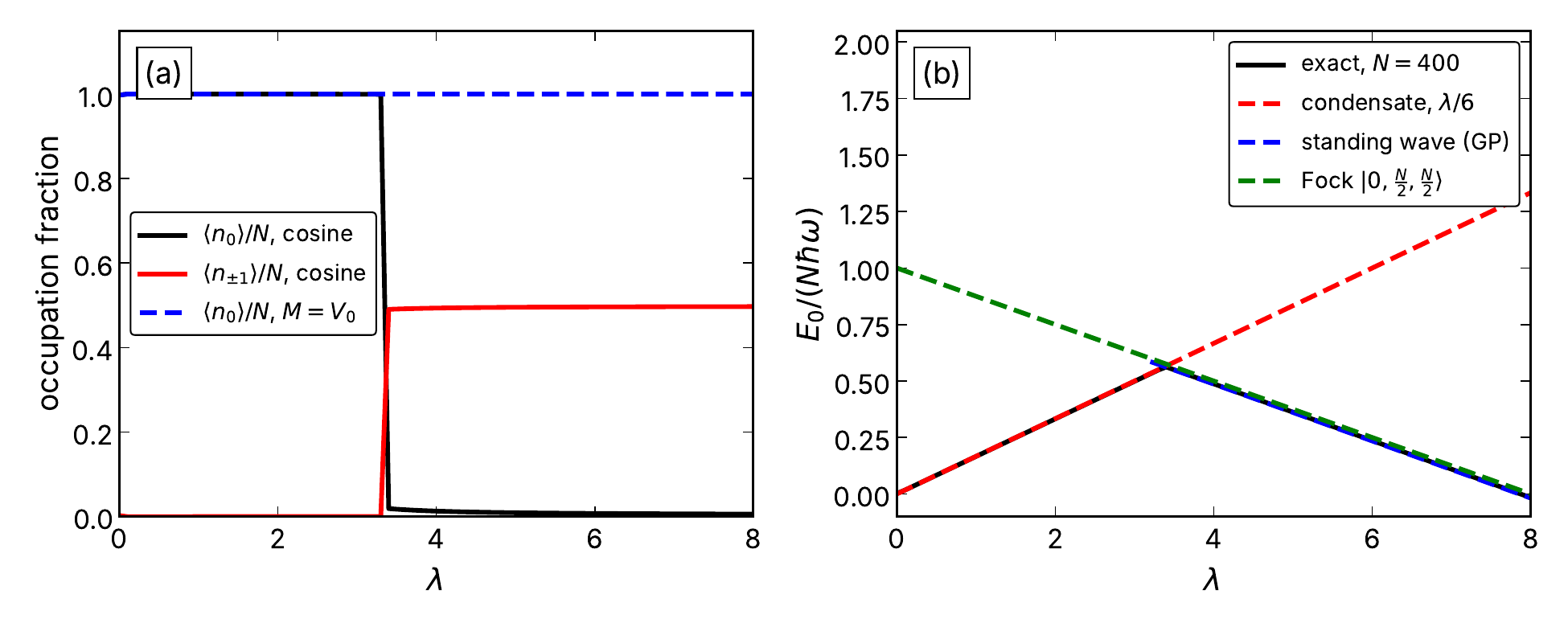}
\caption{Ground state of Eq.~(\ref{eq:H}) for $N=400$ and $c=0$.
(a) Occupation fractions $\avg{n_0}/N$ (black) and $\avg{n_{\pm1}}/N$ (red) versus coupling, and $\avg{n_0}/N$ for a momentum-independent vertex $M=V_0=1$ (blue dashed).
(b) Ground-state energy per particle: exact diagonalisation (black), the $q=0$ condensate (red dashed), the minimised GP standing wave of Eq.~(\ref{eq:GP}) (blue dashed), and the Fock state $\ket{0,N/2,N/2}$ (green dashed).}
\label{fig:transition}
\end{figure}

Figure~\ref{fig:transition}(a) shows the ground-state occupations for $N=400$.
The condensate fraction $\avg{n_0}/N$ stays above $0.99$ up to $\lambda_c$ and then jumps to about $0.02$, while $\avg{n_{+1}}=\avg{n_{-1}}$ jump from zero to $0.49$.
For even $N$ below $\lambda_c$ the ground state has $P=0$; above $\lambda_c$ the three momentum sectors become quasi-degenerate (Sec.~\ref{sec:nature}).
With the momentum-independent vertex the condensate remains in $q=0$ at all couplings.

Three observations identify the transition as first order.
First, the occupations are discontinuous already at moderate $N$ and the jump sharpens with $N$.
Second, the GP functional, Eq.~(\ref{eq:GP}), has two coexisting minima near the transition.
Writing $\psi_0=\sqrt{1-2d^2}$, $\psi_{\pm1}=d\,e^{i\theta_\pm}$, the energy is $e=\lambda/6+d^2[2-\lambda/2+(\lambda/6)(1+\cos(\theta_++\theta_-))]+O(d^4)$; the pair-conversion term is minimised by $\theta_++\theta_-=\pi$, so the curvature is $2(1-\lambda/4)$ and the uniform condensate is metastable for $\lambda_c<\lambda<\lambda_s=4$ (the same instability follows from the Bogoliubov frequency $\omega=\sqrt{(1-\lambda/6)^2-(\lambda/12)^2}$).
Third, the minimum gap between the two lowest states of the $P=0$ block at the avoided crossing decreases faster than exponentially: $1.11\times10^{-4}$, $1.88\times10^{-6}$, $1.59\times10^{-8}$ and $6.8\times10^{-11}$ for $N=20$, 30, 40 and 50, the logarithm dropping by $4.1$, $4.8$ and $5.5$ per ten particles.
This behaviour contrasts with the continuous crossover found for attractive bosons on a ring~\cite{Kanamoto2003}, where the mean-field transition is of second order.

Figure~\ref{fig:transition}(b) compares the exact ground-state energy with three variational energies.
The condensate gives $e_c=\lambda U_0/24=\lambda/6$.
The Fock state $\ket{0,N/2,N/2}$ gives $e_f=1+\lambda U_1/16=1-\lambda/8$ for $N\to\infty$.
The minimised GP standing wave lies slightly below $e_f$, by $0.0142$ at $\lambda=5$, and coincides with the exact energy for $N=400$ within the line width.

\section{Large-$N$ limit}\label{sec:largeN}

\begin{figure}[t]
\centering
\includegraphics[width=\textwidth]{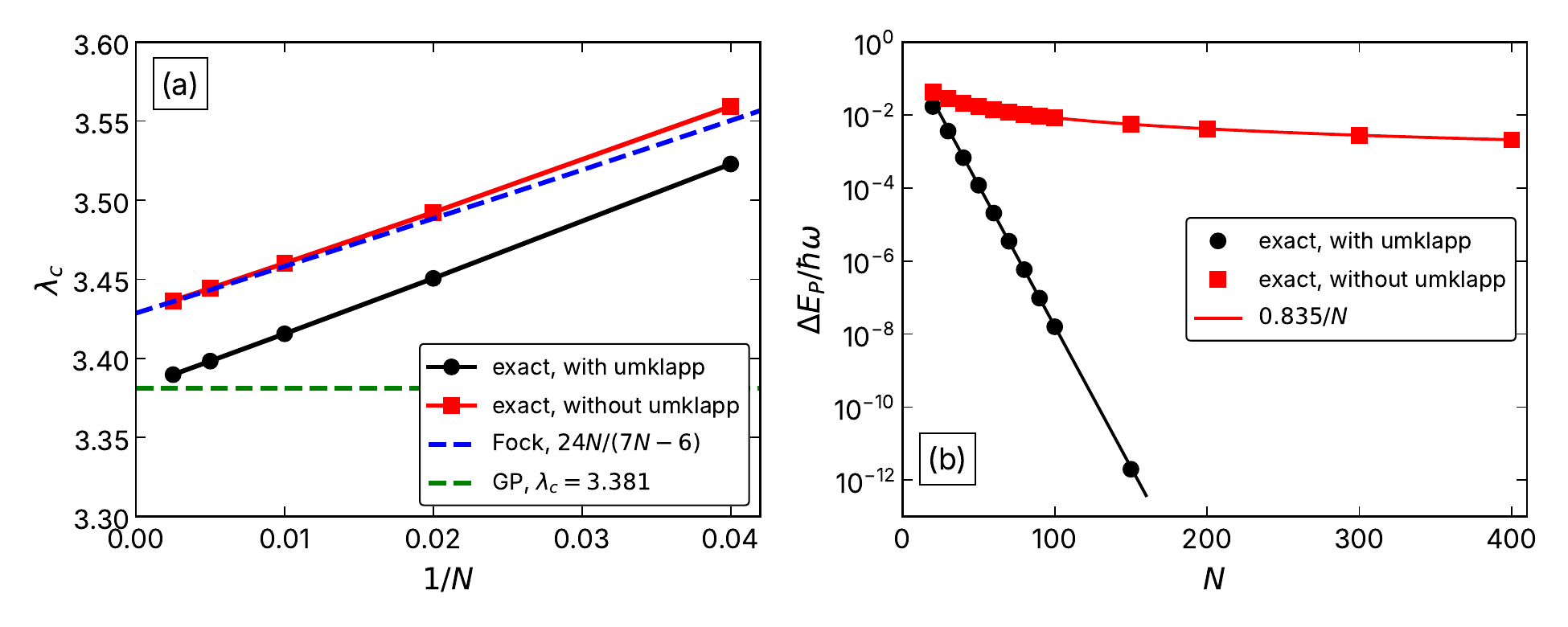}
\caption{(a) Critical coupling versus $1/N$ from exact diagonalisation with umklapp (black circles) and without umklapp (red squares), compared with the Fock-state crossing, Eq.~(\ref{eq:lcFock}) (blue dashed), and the GP critical coupling $\lambda_c^{\rm GP}=3.381$ (green dashed).
(b) Splitting $\Delta E_P$ between the lowest states of the three momentum sectors $P=0,1,2$ at $\lambda=5$, with umklapp (black circles; line: $\propto e^{-0.178N}$, fitted for $40\le N\le100$; for $N\gtrsim150$ the splitting is below the solver tolerance) and without umklapp (red squares; line: $0.835/N$).}
\label{fig:scaling}
\end{figure}

Equating the condensate energy with the minimum of Eq.~(\ref{eq:GP}) on the standing-wave branch gives $\lambda_c^{\rm GP}=3.3812$.
Figure~\ref{fig:scaling}(a) shows the exact critical couplings for $N=25$ to 400.
They approach $\lambda_c^{\rm GP}$ as $1/N$: the differences are $0.0694$, $0.0343$, $0.0171$ and $0.0085$ for $N=50$, 100, 200 and 400, i.e.\ $\lambda_c(N)\simeq\lambda_c^{\rm GP}+3.4/N$.
The GP minimum also reproduces the extensive part of the ground-state energy: at $\lambda=5$ the exact $(E_0-E_f)/N$ is $-0.01254$, $-0.01333$, $-0.01374$ and $-0.01395$ for $N=50$ to 400 and extrapolates in $1/N$ to $-0.01416$, equal to the GP value; the $q=0$ admixture is $0.0092$ (exact, $N=400$) against $0.0091$ (GP).
We have not attempted to derive the coefficient of the $1/N$ correction to $\lambda_c$, which would follow from the Bogoliubov zero-point energies of the two phases~\cite{LNSS2015}.

Without umklapp, the exact critical couplings follow the Fock-state crossing,
\begin{equation}
\lambda_c^{\rm Fock}=\frac{48N}{2U_0(N-1)-U_1(3N-2)}=\frac{24N}{7N-6}\quad(c=0),
\label{eq:lcFock}
\end{equation}
and approach $24/7\simeq3.429$; for $N=400$ the exact value is $3.4364$ against $3.4359$ from Eq.~(\ref{eq:lcFock}), and a $1/N$ extrapolation of the exact values gives $3.4285$.
In the no-umklapp GP functional the $q=0$ admixture of the standing wave vanishes near $\lambda_c$, because its quadratic coefficient, $-1+\lambda(2U_{01}-3U_1)/24$, is positive there; the Fock crossing is then the GP result.
The difference between $24/7$ and $3.381$ is thus entirely due to umklapp scattering.

\section{Fragmented state or symmetry-restored condensate?}\label{sec:nature}

Above $\lambda_c$ the ground state has two eigenvalues of the one-body density matrix close to $N/2$, so it is fragmented in the Penrose-Onsager sense.
To leading order in $N$ its energy is that of a single condensate in a standing-wave orbital, $\psi_0+\psi_{+1}e^{i\theta}+\psi_{-1}e^{-i\theta}$ with $|\psi_{\pm1}|^2\simeq1/2$ (Sec.~\ref{sec:largeN}).
These two statements are compatible, because a symmetric eigenstate obtained by projecting a symmetry-breaking condensate onto a symmetry sector is fragmented, while its energy differs from that of the condensate only at order $N^0$~\cite{Mueller2006}.
The useful question is therefore which symmetry the standing wave breaks.
The answer differs between the two versions of the model, and it is visible in the splitting $\Delta E_P$ of the lowest states in the three momentum sectors [Fig.~\ref{fig:scaling}(b)].

Without umklapp, $L=n_{+1}-n_{-1}$ is conserved, all values of the phase $\theta$ are degenerate, and the standing wave breaks a continuous symmetry.
Its symmetric eigenstates form a rotor tower in $L$, the finite-size counterpart of a continuously broken symmetry~\cite{Anderson1952}.
For the Fock states $\ket{0,\tfrac{N+L}2,\tfrac{N-L}2}$ the exchange term gives $E_L-E_0=\lambda|U_1|L^2/(48N)$, i.e.\ a moment of inertia $I=24N/(\lambda|U_1|)$.
For even $N$, odd $L$ requires an odd occupation of the $q=0$ mode, which costs energy of order one, so the lowest states in the $P=1,2$ sectors have $L=\mp2$ (we checked this explicitly for $N=60$).
Their splitting from the ground state is then predicted to be $4\lambda|U_1|/(48N)=0.833/N$ at $\lambda=5$, and the exact splitting follows $0.835/N$ over the whole range $20\le N\le400$.
Each member of the tower has a sharp $L$; the ground state has $L=0$ and no fluctuations of $n_{+1}-n_{-1}$ at all, i.e.\ it is a fragmented Fock state of the type discussed by Nozi\`eres, dressed only by the pair conversions $\ad_0\ad_0a_{+1}a_{-1}$.

With umklapp, $L$ is conserved only modulo 3.
The umklapp terms make the GP energy depend on $\theta$ and pin it to three equivalent values, so the standing wave breaks the discrete $\mathbb{Z}_3$ translation symmetry.
The three momentum sectors then become degenerate exponentially in $N$, $\Delta E_P\propto e^{-0.178N}$, as for tunnelling between three macroscopically distinct configurations.
The pinning comes from the umklapp terms linear in $\psi_0$, of order $U_{\rm um}\psi_0$, and therefore vanishes at $c=1/2$, where $U_{\rm um}=0$ and the two versions of the model coincide.
Consistently, the relative occupation now fluctuates: the standard deviation of $n_{+1}-n_{-1}$ at $\lambda=5$ is $2.7$, $4.2$, $6.0$, $8.6$ and $12.2$ for $N=25$ to 400, i.e.\ $\simeq0.61\sqrt N$ for large $N$.
This is number-squeezed relative to a coherent state (variance reduced by a factor of about $2.7$), but it grows with $N$, unlike the sharp Fock state of the no-umklapp model.
The exact eigenstate is thus the momentum-projected version of a pinned standing-wave condensate, a superposition of three macroscopically distinct configurations~\cite{Cirac1998}, analogous to the bright soliton of attractive bosons on a ring~\cite{Kanamoto2003}.

In both cases the fragmentation is protected only by the symmetry.
A perturbation that breaks the conservation of $L$ and exceeds the tower splitting selects a single standing wave~\cite{Mueller2006}; this scale is $1/N$ without umklapp and exponentially small with umklapp.

\section{Mechanisms}\label{sec:mech}

\subsection{Exchange}
Consider the Fock states $\ket{N,0,0}$ (condensate), $\ket{0,N,0}$ (single moving condensate) and $\ket{0,\tfrac N2,\tfrac N2}$.
Their diagonal energies follow from the channel constants of Eq.~(\ref{eq:channels}):
\begin{equation}
E_c=\frac{\lambda U_0(N-1)}{24},\qquad
E_s=N+\frac{\lambda U_1(N-1)}{24},\qquad
E_f=N+\frac{\lambda U_1(3N-2)}{48}.
\end{equation}
In $E_f$ the direct terms within each $\pm1$ mode contribute $2\cdot\frac N2(\frac N2-1)$ pairs and the cross terms between $+1$ and $-1$ four orderings (two direct, two exchange) of $\frac N2\cdot\frac N2$ pairs.
Hence
\begin{equation}
E_s-E_f=-\frac{\lambda U_1N}{48},
\label{eq:exchange}
\end{equation}
which is entirely exchange energy.
In the standing-wave language of Sec.~\ref{sec:nature} the same quantity is the interference energy of the density modulation: a coherent state with $|\psi_{\pm1}|^2=1/2$ has the energy $E_f$ at order $N$, because no term of Eq.~(\ref{eq:H}) within the $\pm1$ sector depends on the relative phase.
For repulsive $U_1>0$ the single moving condensate is lower, which is Nozi\`eres' result.
For attractive $U_1<0$ the same exchange term favours the fragmented state.
Fragmentation is therefore predicted for $U_1<0$, i.e.\ $c<2$.

Figure~\ref{fig:mech}(a) tests this prediction at $\lambda=8$, well above the transition, and $N=100$.
Without umklapp, $F$ drops from 1 to 0 at $c^*=2.02$, and for $c>2$ the ground state is the single moving condensate with $\avg{n_{+1}}=N$ or $\avg{n_{-1}}=N$, in agreement with Eq.~(\ref{eq:exchange}).
The momentum-independent vertex has $U_0=U_1$, so $E_c<E_s$ for all $\lambda$ and the condensate never leaves $q=0$, consistent with Fig.~\ref{fig:transition}(a).

\subsection{Umklapp}
\begin{figure}[t]
\centering
\includegraphics[width=\textwidth]{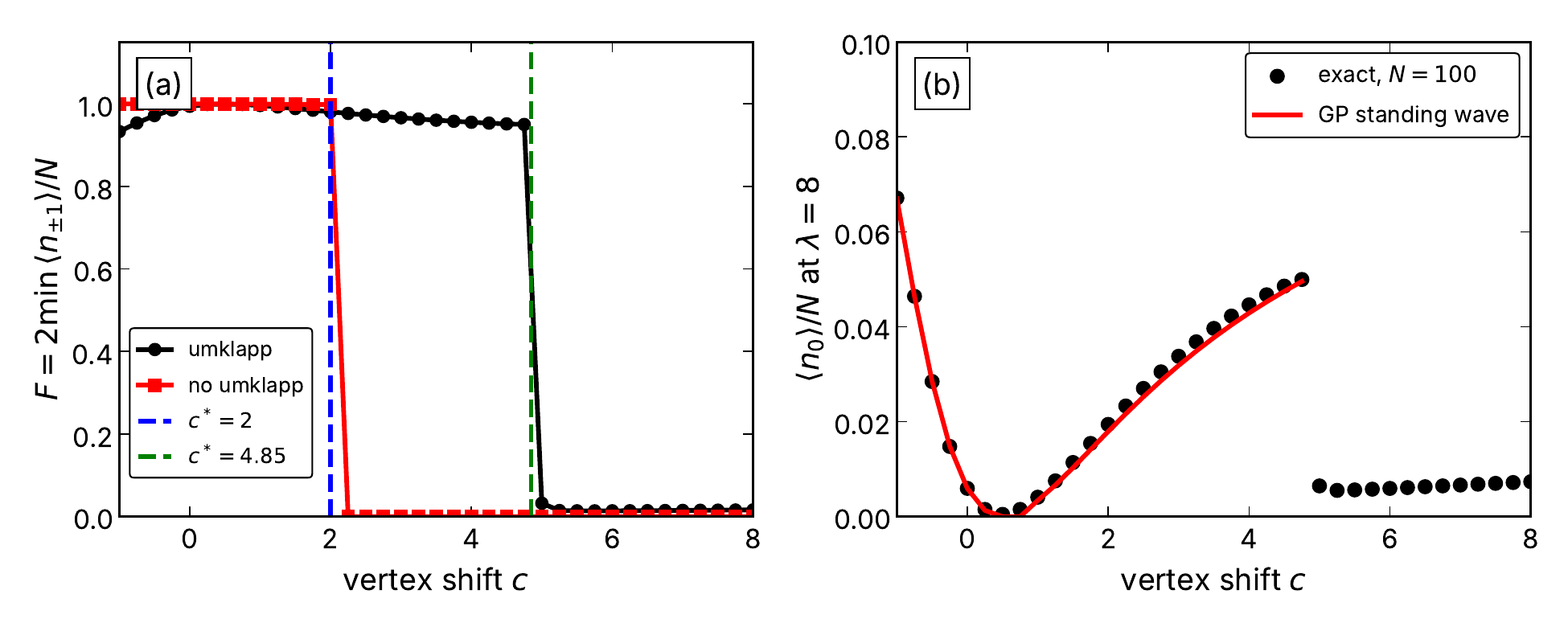}
\caption{Test of the mechanisms by a constant shift $c$ of the vertex, Eq.~(\ref{eq:M}), at $\lambda=8$ and $N=100$.
(a) Fragmentation $F=2\min\avg{n_{\pm1}}/N$ with umklapp (black) and without umklapp (red), with the exchange boundary $c^*=2$ (blue dashed) and the GP boundary with umklapp, $c^*=4.85$ (green dashed).
(b) $q=0$ admixture $\avg{n_0}/N$ of the exact ground state (symbols) and of the minimised GP standing wave (red).}
\label{fig:mech}
\end{figure}

With umklapp the $\pm1$ state survives well beyond $c=2$, up to $c^*=4.96$ for $N=100$ and $4.90$ for $N=200$, approaching the GP boundary $c^*_{\rm GP}=4.85$ [Fig.~\ref{fig:mech}(a)].
The reason is the coherent coupling of the standing wave to the $q=0$ mode.
The umklapp channel contains terms linear in $\psi_0$, such as $\psi_0^*\psi_{+1}^*\psi_{-1}\psi_{-1}$, which exist only when both $\pm1$ modes are occupied.
A small coherent $q=0$ amplitude then lowers the standing-wave energy by an amount of order $U_{\rm um}^2$, which has no counterpart for the single moving condensate. In the real-space language, the single moving condensate carries a persistent current around the triangle, whereas the standing wave does not.
For $c>2$ this gain outweighs the exchange penalty of Eq.~(\ref{eq:exchange}) until $c\simeq4.85$.

Figure~\ref{fig:mech}(b) shows the corresponding fingerprint.
The $q=0$ admixture of the exact ground state follows the GP prediction across the whole range and vanishes near $c=1/2$, where $U_{\rm um}=0$.
For $c>c^*$ the admixture drops to the small value of the single moving condensate.

\section{Conclusions}\label{sec:conc}

On a three-site Bose ring, density-assisted hopping of the opposite sign to the single-particle hopping inverts the band and drives a first-order transition from the $k=0$ condensate into the degenerate pair $k=\pm2\pi/3$.
At leading order in $N$ the resulting state has the energy of a GP standing wave, as expected for a few-mode system with $1/N$ interactions, and the GP functional gives its critical coupling and $q=0$ admixture.
What the exact solution adds is the nature of the many-body state, which depends on the symmetry the standing wave breaks.
When momentum is conserved as an integer, the standing-wave phase is a continuous zero mode, the eigenstates form a rotor tower with $1/N$ splitting given by the exchange energy, and the ground state is a fragmented Fock state precisely when the interaction among the competing modes is attractive, as Nozi\`eres' argument predicts once the sign of the interaction is reversed; otherwise a single current-carrying condensate forms.
Umklapp scattering pins the phase, makes the momentum sectors degenerate exponentially in $N$, and through coherent coupling to the $q=0$ mode stabilises the standing wave over a wider range; its Penrose-Onsager fragmentation is then a property of the exactly symmetric eigenstate only, and is destroyed by exponentially small symmetry-breaking perturbations.
The model thus allows the distinction between fragmentation and symmetry-restored condensation, and the role of exchange in both, to be followed quantitatively in a numerically exact calculation.

\section*{Acknowledgements}
The model goes back to the author's M.Sc.\ thesis at Bilkent University; the supervision of Tu\u{g}rul Hakio\u{g}lu is gratefully acknowledged.
Claude, ChatGPT, Jenni AI and Grammarly were used for calculations, figure design and English-language improvement.

\section*{Data availability}
All codes and data used to produce the results and figures of this paper (Python with NumPy and SciPy) are provided as Supplemental Material.

\appendix
\section{Integer encoding of the Fock basis}\label{app:encoding}
A Fock state $\ket{n_0,n_1,\dots}$ with occupations below $K_l$ in mode $l$ is stored as the integer $c=\sum_l n_l s_l$, $s_l=\prod_{j<l}K_j$.
Creation and annihilation operators act as $c\mapsto c\pm s_l$ with factors $\sqrt{n_l+1}$ and $\sqrt{n_l}$; a creation operator acting on a full digit must return zero explicitly, since the addition would otherwise carry into the next digit.
Each term of the Hamiltonian then maps a basis integer to at most one other integer, and conserved quantities that are linear in the occupations ($N$, $P$, $L$) are read off the digits to restrict the basis to one symmetry block.
This is the standard integer representation of exact-diagonalisation codes~\cite{Lin1990,Sandvik2010,ZhangDong2010,QuSpin2017}; it was formulated independently, as a bijective base-$K$ numeral, in Ref.~\cite{thesis}.
With per-mode bases it accommodates spins ($K=2S+1$) and fermions ($K=2$, with the Jordan-Wigner sign given by the number of occupied modes below $l$).
As checks, the matrix of Eq.~(\ref{eq:H}) agrees with an independent Kronecker-product construction to $10^{-14}$, block spectra reproduce full spectra to $10^{-12}$, and the same code reproduces the ground-state energies of periodic spin-$\tfrac12$ Heisenberg rings ($L\le16$) to $10^{-9}$ and the Hubbard dimer exactly.

\end{document}